# Channel Interaction and Current Level Affect Across-Electrode Integration of Interaural Time Differences in Bilateral Cochlear-Implant Listeners[1]

Katharina Egger, Piotr Majdak[a], and Bernhard Laback

Acoustics Research Institute
Austrian Academy of Sciences
Wohllebengasse 12-14
A-1040 Vienna, Austria

Email addresses: katharina.egger@oeaw.ac.at (Katharina Egger), piotr@majdak.com (Piotr Majdak), bernhard.laback@oeaw.ac.at (Bernhard Laback)

**Running title:** Across-Electrode Integration

File: ITDMultEl 3E.odt; August 3, 2015

---

[a] Corresponding author: Acoustics Research Institute, Austrian Academy of Sciences, Wohllebengasse 12-14, A-1040 Vienna, Austria. Telephone: +43 1 51581-2511. Email: piotr@majdak.com

# ABSTRACT

Sensitivity to interaural time differences (ITDs) is important for sound localization. Normal-hearing listeners benefit from across-frequency processing, as seen with improved ITD thresholds when consistent ITD cues are presented over a range of frequency channels compared to when ITD information is only presented in a single frequency channel. This study aimed to clarify whether cochlear-implant (CI) listeners can make use of similar processing when being stimulated with multiple interaural electrode pairs transmitting consistent ITD information. ITD thresholds for unmodulated, 100-pulse-per-second pulse trains were measured in seven bilateral CI listeners using research interfaces. Consistent ITDs were presented at either one or two electrode pairs at different current levels, allowing for comparisons at either constant level per component electrode or equal overall loudness. Different tonotopic distances between the pairs were tested in order to clarify the potential influence of channel interaction. Comparison of ITD thresholds between double pairs and the respective single pairs revealed systematic effects of tonotopic separation and current level. At constant levels, performance with double-pair stimulation improved compared to single-pair stimulation, but only for large tonotopic separation. Comparisons at equal overall loudness revealed no benefit from presenting ITD information at two electrode pairs for any tonotopic spacing. Irrespective of electrode-pair configuration, ITD sensitivity improved with increasing current level. Hence, the improved ITD sensitivity for double pairs found for a large tonotopic separation and constant current levels seems to be due to increased loudness. The overall data suggest that CI listeners can benefit from combining consistent ITD information across multiple electrodes, provided sufficient stimulus levels and that stimulating electrode pairs are widely spaced.

# INTRODUCTION

Binaural hearing is important for the localization and segregation of sound sources. While bilateral cochlear implantation has been shown to enable some basic left versus right localization, cochlear-implant (CI) listeners' abilities to localize sounds and to understand speech in background noise are still limited compared to normal-hearing listeners (e.g., Kerber and Seeber 2012; Majdak et al. 2011; Schleich et al. 2004). It has been shown that the performance of bilateral CI listeners in sound localization is mainly mediated by the perception of interaural level differences (Grantham et al. 2007; Grantham et al. 2008; Seeber and Fastl 2008). It can be assumed that the limitation in performance is linked to the restricted access to interaural time difference (ITD) information with current envelope-based bilateral CI systems and the resulting poor sensitivity to ITD cues of CI listeners (see Laback et al. 2015, for a recent review).

There is a body of literature showing that CI listeners are most sensitive to ecologically relevant ITD cues when stimulating with a single, pitch-matched, interaural electrode pair by the use of direct, interaurally coordinated stimulation (e.g., Laback et al. 2007; Litovsky et al. 2010; Majdak et al. 2006; van Hoesel 2007; van Hoesel et al. 2009). CI listeners' sensitivity to ITD cues under more realistic conditions, i.e., with stimulation at multiple-electrode pairs, is a new and developing field of research with only a few studies published yet (e.g., Francart et al. 2015; Ihlefeld et al. 2014). In normal hearing, ITD sensitivity has been found to improve when the same ITD information is presented across different frequency bands compared to when the ITD is only presented in a single frequency channel (e.g., Buell and Hafter 1991; Buell and Trahiotis 1993). This beneficial use of combined information across frequencies is further referred to as across-frequency integration of consistent ITD information. Accordingly, it can be expected that the sensitivity to ITD improves with an increasing number of stimulating electrode pairs carrying consistent ITD. However, electrical current spread within the cochlea can cause overlapping excitation patterns when multiple electrodes are stimulated (e.g., Chatterjee et al. 2006). Such interactions between neighboring channels might deteriorate transmitted timing cues and, thus, the sensitivity to ITD cues presented at multiple-electrode pairs. Further, with increasing number of pairs, the perceived loudness may change due to loudness summation (e.g., McKay et al. 2001). When fitting clinical CI processors, level adjustments are necessary to compensate for multi-channel loudness summation: For multiple-electrode stimulation, the current levels of the individual component electrodes are usually lower compared to when a single electrode is evaluated at the same loudness. Such level adjustments may affect the ITD sensitivity, as observed in normal-hearing listeners who showed deteriorating performance with decreasing level (Dietz et al. 2013; Hershkowitz and Durlach 1969; Zwislocki and Feldman 1956). Some evidence for a similar effect of level on ITD sensitivity in CI listeners was shown for two CI listeners (van Hoesel 2007).

Recent studies have focused on ITD perception of amplitude-modulated, high-rate pulse trains when consistent ITD cues are presented at multiple-electrode pairs representing the location of a single sound source (Francart et al. 2015; Ihlefeld et al. 2014). Ihlefeld et al. (2014) presented consistent ITD cues at two pitch-matched, interaural electrode pairs (i.e., a double pair) to investigate the combination of temporally interleaved envelope-ITD information across different stimulation sites. The two electrode pairs were widely spread along the electrode array. The current levels of the component electrodes were kept constant for single- and double-pair-stimulation. Ihlefeld et al. found slightly and consistently better ITD sensitivity for the double pair compared to single pairs (see their statistical results from p. 5). Francart et al. (2015) measured ITD sensitivity for single- and three-electrode pair stimuli that were loudness-balanced across conditions. For the latter, either a tonotopic separation of four electrodes or adjacent electrodes were used. Francart et al. did not find any significant difference in performance of single-pair versus triple-pair stimulation when the envelope modulations were kept synchronous across electrodes. However, ITD sensitivity was found to deteriorate with increasing across-channel envelope asynchrony, particularly for the triple pair with adjacent electrodes. An analysis of the across-channel excitation patterns suggested that the decrease in ITD sensitivity was most likely due to a reduction in dead time between subsequent pulses. No significant effect of tonotopic separation was found, irrespective of channel envelope (a)synchrony.

The present work aims to elucidate the perception of consistent ITD cues presented at multiple-electrode pairs. We focused on the perception of ITDs contained in unmodulated, low-rate pulse trains, where only preliminary investigations have been conducted (e.g., Jones et al. 2009; Jones et al. 2013). In the present study, ITD sensitivity was investigated in regard to three hypotheses. First, it was hypothesized that ITD sensitivity improves with an increasing number of stimulating electrode pairs carrying consistent ITD. Further, it was hypothesized that the larger the tonotopic distance between the stimulating electrode pairs, the better the ITD sensitivity. A larger tonotopic distance reduces the possibility of overlapping excitation patterns across electrodes which might degrade transmitted timing cues. Finally, to verify whether level adjustments (necessary to compensate for multi-channel loudness summation) affect ITD sensitivity, it was hypothesized that ITD sensitivity improves with increasing current level.

To clarify the potential contribution of each of the described factors, ITD thresholds for unmodulated, 100-pulse-per-second (pps) pulse trains were measured by stimulating either one or two interaural electrode pairs at different current levels allowing for comparisons at either constant levels or equal overall loudness. For the comparison at constant levels, the current levels of the component electrodes were the same for both single and double pairs, yielding increased overall loudness for the double pair compared to the respective single pairs. For the comparison at equal overall loudness, the double pair had lowered current levels compared to the respective single pairs yielding equal loudness for single and double pairs. Different tonotopic distances between the pairs were tested.

**METHODS**

*Listeners*

Seven listeners (three female, four male) bilaterally supplied with 12-electrode CIs (manufactured by MED-EL, Austria) participated in the experiment. The implants provide monopolar stimulation with an extra-cochlear ground electrode. All listeners had good speech perception in daily communication. Individual listener data are shown in Table 1. Three listeners (CI1, CI24, and CI52) had previous experience with psychoacoustical ITD experiments. All listeners were paid an hourly wage for their participation.

*Stimuli and apparatus*

Unmodulated, 300-ms, biphasic pulse trains at a rate of 100 pps were used. Each phase of a pulse had a duration of 26.7 µs by default. For some listeners, this duration had to be increased for certain electrodes since a sufficient maximum comfortable level (MCL) could not be achieved in the fitting procedure. Any deviations from the standard phase duration are listed in Table 2. The interphase gap was 0 µs for C40+ implants and 2.1 µs for Pulsar and Concerto implants. The pulse trains were presented at either one or two interaurally pitch-matched electrode pairs simultaneously, which were selected in the pretests (see Sec. Electrode-fitting and pitch-matching). The selected electrode pairs were labeled from A (the most apical pair) to D (the most basal pair) (Fig. 1). The pairs B, C, and D were used in the main experiment, pair A was only used for ITD training (see Sec. ITD sensitivity training). In case of double pairs, i.e., when stimulating two electrode pairs, both pairs had the same ITD and the temporal offset between the interleaved pulses across component electrodes corresponded to half the interpulse interval (Fig. 2). The first pulse of the pulse train was always presented to the more apical electrode. Two double pairs were tested: BD represented a large tonotopic separation with a distance between the pairs of 14 mm on average, and CD represented a small tonotopic separation with a distance of 6 mm on average. The additional condition DD represented a double pair with infinitely narrow electrode spacing and was achieved by stimulating the pair D with double rate, i.e., 200 pps. The electrodes used in each single pair and the tonotopic distances employed in the double pairs are listed in Table 3.

The stimuli were presented at high, middle, or low current levels denoted by $L^{HI}$, $L^{MI}$, and $L^{LO}$, respectively. Single and double pairs for a specific level condition are hereafter denoted by superscripts according to the respective level condition. The three level conditions $L^{HI}$, $L^{MI}$, and $L^{LO}$ were successively determined in the following steps: (1) For condition $L^{HI}$, the levels of the single pairs ($B^{HI}$, $C^{HI}$, $D^{HI}$) were matched at an equal, comfortable loudness. (2) For the double pairs ($BD^{HI}$, $CD^{HI}$), the respective component-electrode levels were kept constant which resulted in an increased overall loudness. If necessary, the levels were adjusted to elicit a centered auditory image. (3) For the condition $L^{MI}$, the double pairs ($BD^{MI}$, $CD^{MI}$) were matched in overall loudness to the respective single pairs of conditions $L^{HI}$ to accommodate the effect of loudness summation in multiple-electrode stimulation. (4) The component-electrode levels of the single pairs ($B^{MI}$, $C^{MI}$, $D^{MI}$) were the same as those of the double pairs ($BD^{MI}$, $CD^{MI}$). (5) In condition $L^{LO}$, the levels of the double pairs ($BD^{LO}$, $CD^{LO}$) corresponded to 85 % of the levels of the double pairs in condition $L^{MI}$, and (6) the component-electrode levels of the single pairs were the same as those of the double pairs. For more details, see Sec. Adaptive loudness matching, single pairs and Sec. Current level determination for double pairs.

All stimuli were generated on a personal computer and then controlled and sent to the CIs via a research interface box (RIB2, developed at the Institute of Ion Physics and Applied Physics, Leopold-Franzens-University of Innsbruck, Austria). The RIB2 allows direct and interaurally coordinated stimulation of two CIs.

*Pretests*

First, electrode-fitting and pitch-matching procedures were performed to determine up to four interaural electrode pairs which elicited the same sensation of pitch. Second, an adaptive loudness matching procedure was conducted so that stimulation of the different single pairs elicited approximately the same loudness. Third, lateralization discrimination tests were performed in order to ensure that the listeners were sensitive to ITDs at each of the selected single pairs. During those tests, feedback was provided for the purpose of training, which was required especially for the listeners who had no prior experience with psychoacoustical experiments measuring ITD sensitivity. Last, the current levels for the double pairs were determined.

***Electrode-fitting and pitch-matching***

For the pitch matching, the stimuli were unmodulated, 300-ms, pulse trains at a rate of 1515 pps. The high pulse rate was used in order to reduce the confounding effect of rate pitch which is only salient at low rates. First, the electric dynamic range (DR), determined by the threshold (THR) and the MCL, and the comfortable level (CL) of each electrode in both ears were manually determined using a continuous loudness scale. Subsequently, a pitch-matching procedure (as described previously, e.g., in Majdak et al. 2006) was performed to determine interaurally pitch-matched electrode pairs. Briefly, a monaural pitch magnitude estimation procedure based on the CLs was used in order to reduce the number of pitch-matched, interaural electrode pair candidates. For each electrode pair candidate, an iterative, interaural loudness balancing procedure was performed repeatedly and in random order. In a final step, pitch discriminability for each electrode pair candidate was measured in a pitch ranking procedure and four pitch-matched pairs (A to D) were selected. These single pairs were aimed to be roughly evenly distributed along the tonotopic range of the electrodes ranging from 1 to 12 (numbered in ascending order from apical to basal regions in the cochlea, Fig. 1). Note that, because of medical reasons, 5 out of 12 electrodes in the left implant of listener CI52 were in-active at the time of testing. Therefore, only three pitch-matched electrode pairs (B to D) were identified for CI52.

For all subsequent procedures, unmodulated, 300-ms, pulse trains at a rate of 100 pps were used. The electric DR (i.e., the difference between THR and MCL) and CL of the electrodes selected in the pitch-matching procedure were manually determined for each ear using the continuous loudness scale. In order to verify that, for an ITD of zero, the stimulation of the selected pitch-matched single pairs evoked a centered auditory image at a comfortable loudness, the stimuli were presented repeatedly at each single pair separately. In a manual procedure (controlled by the experimenter), the listeners were alternately asked to judge the loudness and to indicate the perceived auditory image position on a left-center-right scale. The levels were adjusted so that the stimuli were perceived at a comfortable loudness (labeled as "middle" on the loudness scale). Reproducibly perceived position deviations from the center were compensated by small level adjustments.

*Adaptive loudness matching, single pairs*

An adaptive loudness matching procedure was used to iteratively match the single pairs in loudness against each other (Jesteadt 1980). In that task, one single pair was defined as reference which was stimulated with a fixed current level, and another single pair was defined as target which was stimulated with a variable current level. Note that level here refers to the left and right ear levels of the binaural stimulus.

For each listener, one reference pair was selected to which the loudness of the remaining single pairs was matched. Loudness matches were obtained in an adaptive two-interval, two-alternative forced-choice procedure using two randomly interleaved sequences with different decision rules converging at the point of equal loudness of reference and target. The adaptive level adjustment for the target followed a two-down, one-up rule converging at 71 % of louder-perceived target stimuli (upper sequence) and a one-down, two-up rule converging at 29 % of louder-perceived target stim-uli (lower sequence, Levitt 1971). Each trial consisted of two intervals separated by a 300-ms pause. One randomly chosen interval contained the reference, whose level was set to the corresponding CL. The other interval contained the target. The listeners had to indicate the louder interval by pressing the corresponding button on the response pad. At the beginning of each sequence, the target level was set to 20 % DR above and 20 % DR below the CL of the corresponding single pair for the upper and lower sequence, respectively. When it was observed that listeners had difficulties with the task, initial levels were further increased (upper sequence) and decreased (lower sequence) in order to increase the contrast between reference and target at the beginning of the sequences and, hence, to simplify the task. In a few cases, the initial level had to be reduced to not exceed the MCLs. For each sequence, the initial step size was 10 % DR. After each reversal, that step size was decreased by a factor of 0.7 until it reached the minimum step size of 2 % DR. For the remaining reversals, the step size was kept constant. Each sequence was finished after twelve reversals and the levels from the last six reversals were averaged to estimate the 71 % (upper sequence) or 29 % (lower sequence) louder levels for that run. The level representing the loudness match of reference and target was then calculated as the arithmetic average of the 71 % and 29 % levels. In cases where there was a lack of convergence of one of the individual sequences, both upper and lower sequences were repeated for that run. Each listener performed at least two valid runs for each target, except CI61, who performed only one run for matching the pairs C and D, and CI52, who had difficulties matching C and D. Since CI52 was available only for a limited amount of time, his/her loudness matches were performed manually by presenting reference and target alternately and in random order. Listener CI52 was asked whether the second stimulus was perceived louder, softer, or equally loud compared to the first stimulus; the experimenter adjusted the level of the target respectively. This procedure was repeated until the listener reproducibly perceived both reference and target as equally loud.

For each target, the final current level was determined by arithmetically averaging the levels obtained from the runs. Finally, the resulting level for each single pair was verified using the aforementioned left-center-right scale to evaluate the lateral position of the auditory image. Potential deviations from the center were compensated by 1) decreasing the level in that ear, to which the stimuli were perceived and 2) increasing the level in the other ear. All level changes were made in uniform steps of % DR.

*ITD sensitivity training*

ITD sensitivity for the selected pitch-matched single pairs was verified in an adaptive two-interval, two-alternative forced-choice procedure with visual feedback. A trial consisted of two intervals separated by 300-ms pauses. The first interval contained the reference stimulus with zero ITD. The second interval contained the target stimulus with non-zero ITD, i.e., the pulses at one ear were delayed relative to the other ear. The listeners had to indicate to which side (left or right) the second stimulus was perceived compared to the first stimulus by pressing the corresponding button on the response pad. The ITD of the target stimulus was applied randomly either to the left or to the right side. The target ITD was initially set to 800 μs. When the listeners had no prior experience with left versus right discrimination tasks or showed difficulties with the task, the starting ITD was increased up to 1500 μs. The adaptive target ITD adjustment followed a four-down, one-up rule converging at the 84 % point of the psychometric function (Levitt 1971). The initial step size was 500 μs. After each reversal, it was decreased by a factor of 0.7 until it reached the minimum step size of 50 μs. For the remaining reversals, the step size was kept constant. Each run was finished after twelve reversals and the ITDs from the last eight reversals were arithmetically averaged yielding the ITD threshold. Some of the ITD thresholds were measured in blocks of two interleaved adaptive runs.

The ITD training began with the single pair A and continued until the listeners showed stable performance (approximately seven to nine runs). Subsequently, ITD sensitivity at the single pairs B,

C, and D was trained (one or two runs). If the obtained thresholds differed substantially between the pairs, the training of the worse single pair(s) was continued, aiming at comparable thresholds between the pairs B to D.

For CI52, only three pitch-matched electrode pairs were identified. These pairs B, C, and D were trained repeatedly and in random order. The ITD thresholds for B (793 µs, at its best) differed remarkably from the thresholds for C (295 µs) and D (388 µs), despite extensive training with B. Consequently, B was excluded from further measurements. CI60 showed poor performance at A (1074 µs, at its best), despite extensive training. Therefore, the pairs B, C, and D were trained repeatedly and in random order. CI61 showed only poor performance at A (only 25 % of all runs valid; at its best 884 µs – first session, 499 µs – second session) and no sensitivity to ITD at B (no valid runs). Therefore, B was excluded from further measurements and the pairs C and D were trained repeatedly and in random order.

The training was not only conducted in the course of the pretests but also before each test session of the main experiment when it was scheduled on a different day. In that case, listeners were re-familiarized with the task by completing one or two adaptive runs with either A (when it had been identified and had been found to be sensitive to ITD during the pretests) or each of the single pairs used in the main experiment. This way, the training either did not interfere with the tested pairs B, C, and D (for training with A) or affected all tested pairs equally (for training with all pairs).

*Current level determination for double pairs*

In a first step, a manual procedure was used to verify that for an ITD of zero, the stimulation with the double pairs evoked a centered auditory image. In that procedure, the component-electrode levels of each double pair corresponded to those of the respective single pairs (see Sec. Adaptive loudness matching, single pairs). The perceived lateral position was repeatedly evaluated with the aforementioned left-center-right scale and deviations from the center were compensated by 1) decreasing the levels of both electrodes in that ear, to which the stimuli were perceived and 2) increasing the levels of both electrodes in the other ear. All level changes were made in uniform steps of % DR in order to keep the contribution of each individual electrode balanced. The resulting levels represent the level condition $L^{HI}$.

In a next step, the adaptive loudness matching procedure (see Sec. Adaptive loudness matching, single pairs) was used to match the overall loudness of the double pairs to that of the respective single pairs. In that procedure, the double pair was the target and the single pairs were the references. For example, in order to match the double pair CD to either single pair C or single pair D, C was defined as reference 1, D as reference 2, and CD as target. Consequently, the procedure was extended such that the target was compared to two different references in the same block, yielding four randomly interleaved sequences in the same block. The initial current level of the target was set to 20 % DR above and 30 % DR below the level of the corresponding double pair in condition $L^{HI}$, for the upper and lower sequences, respectively. The level representing the loudness match was the average of the levels obtained in a block. Each listener performed at least two valid loudness matching measurements for each double pair, except CI52, whose availability was limited and therefore loudness matches were performed manually (analogous to Sec. Adaptive loudness matching, single pairs).

Then, for each double pair, the perceived lateral position was repeatedly evaluated with the aforementioned left-center-right scale with the resulting levels. When the stimulus was perceived to the left or to the right, the levels of the electrodes in the corresponding ear were slightly decreased, but the levels of the electrodes in the opposite ear were kept constant in order to ensure that the double pair was not perceived louder than the respective single pairs. These level changes were made in uniform steps of % DR across electrodes. The resulting levels represent the second level condition $L^{MI}$. Note that matching the double pairs BD and CD to their respective single pairs yielded two dif-ferent levels for the pair D. In case of potential ambiguity, the corresponding conditions are denoted by a subscript (e.g., $D_B^{HI}$, $D_B^{MI}$, $D_C^{HI}$, and $D_C^{MI}$, but note the difference to BD).

In a final step, current levels of each double pair were further decreased to 85 % of the loudness-matched levels ($BD^{MI}$, $CD^{MI}$) of each component electrode. With the resulting levels, the perceived lateral position was evaluated and deviations from the center were compensated when required. As before, current levels were only decreased and never increased, all in uniform steps of % DR across electrodes. The resulting levels represent the third level condition $L^{LO}$.

*Procedure*

To measure ITD sensitivity, a constant stimuli paradigm using a two-interval, two-alternative forced-choice procedure was employed. The task was the same as in the ITD sensitivity training, only that, this time, the ITD of the target stimulus was not adaptively adjusted but set by the experimenter. Percent correct scores were measured for at least four different ITDs per condition which depended on the individual listeners' sensitivity. The scores were measured in blocks, each of them containing either one particular single or double pair tested at different current levels according to the level conditions. Each block contained between 10 and 30 repetitions per level condition and tested ITD in random order. The blocks were tested in random order yielding a sequence. This sequence was subsequently repeated in reversed order. The number of blocks was such that each ITD was measured at least 60 times, with an equal number of targets to the left and to the right. The listeners took a break after each block and longer blocks also included breaks within one block.

Double pairs were tested for all three level conditions (e.g., $BD^{HI}$, $BD^{MI}$, and $BD^{LO}$). The single pairs were only tested at the two highest current levels (e.g., $B^{HI}$ and $B^{MI}$). In case of single pair D, up to four different levels were used depending on the combination with pair B or pair C ($D_B^{HI}$, $D_B^{MI}$, $D_C^{HI}$, and $D_C^{MI}$). The "double pair" DD (i.e., single pair D stimulated with twice the pulse rate) was tested with the same levels as used for the single pair D (i.e., $DD_B^{HI}$, $DD_B^{MI}$, $DD_C^{HI}$, and $DD_C^{MI}$).

For the listeners CI52 and CI61, only the double pairs CD and DD and the respective single pairs C and D were tested as pair B had been excluded from the measurements within the training. Since time allowed, the thresholds of CI61 were measured for all three level conditions, even for the single pairs. Since listener CI24's time was limited, thresholds for all single pairs and double pair DD were measured only for the highest current level (level condition $L^{HI}$).

For each condition, the percent correct scores were fit with a Weibull function yielding the listener-specific ITD threshold at the 75-% point of the psychometric function.

## RESULTS

Figure 3 shows the ITD thresholds geometrically averaged across listeners for both stimulation types, i.e., single pairs (green bars) and double pairs (gray bars), for large, small, and zero tonotopic separation of the double pairs. Please see the appendix for the individual data. Thresholds in Fig. 3 are shown for the level conditions $L^{HI}$ (top row) and $L^{MI}$ (bottom row). Each panel compares single- and double-pair thresholds for constant levels, i.e., component-electrode levels were kept constant yielding that the double pair was louder than the respective single pairs. For the large tonotopic separation (BD), lower average thresholds were found for the double pair compared to those found for the corresponding single pairs in both level conditions: double-pair thresholds were roughly 20 μs ($L^{HI}$) to 50 μs ($L^{MI}$) lower compared to that of the better single pair. For the small tonotopic separation (CD), average thresholds for single and double pairs were of similar size in both level conditions $L^{HI}$ and $L^{MI}$. For the zero tonotopic separation (DD), far larger average thresholds were found for the double pair than for the corresponding single pair for both current levels tested. Note that for zero tonotopic separation, thresholds are shown for $DD_B$ and $DD_C$, because single pair D and double pair DD were tested at various levels which depended on the combination with B or C (denoted by the subscript *B* or *C*, respectively).

Statistical analysis was performed on the individual ITD thresholds, jointly for the level conditions $L^{HI}$ and $L^{MI}$. Two-way repeated measures analyses of variance (ANOVAs) with the main effects stimulation type (single- vs. double-pair stimulation) and current level ($L^{HI}$ vs. $L^{MI}$) were conducted (MATLAB and Statistics Toolbox Release 2011b) separately for each of the three tonotopic separations. The thresholds were logarithmically transformed in order to fulfill the requirement of homoscedasticity of the data and normality of the residuals. In pretests, the interaction of stimulation type and current level was not significant in all three analyses [$p > 0.4$]. The interaction term was therefore dropped and the ANOVAs were repeated with the main effects only. For the large tonotopic separation, a significant effect of stimulation type [$F_{1,21} = 6.65$, $p = 0.0175$] and current level [$F_{1,21} = 9.39$, $p = 0.0059$] was found: Double-pair thresholds were significantly lower than single-pair thresholds; thresholds for $L^{HI}$ were significantly lower than those for $L^{MI}$. For the small tonotopic separation, a significant effect of current level (significantly lower thresholds for $L^{HI}$ than for $L^{MI}$) [$F_{1,31} = 8.23$, $p = 0.0073$] but no significant effect of stimulation type was found [$F_{1,31} = 0.26$, $p = 0.6122$]. For zero separation, a significant effect of stimulation type [$F_{1,35} = 38.02$, $p < 0.0001$] and current level [$F_{1,35} = 20.11$, $p < 0.0001$] was found: Double-pair thresholds were significantly higher than single-pair thresholds; thresholds for $L^{HI}$ were significantly lower than those for $L^{MI}$.

Figure 4 shows only those ITD thresholds from Fig. 3 which represent equal overall loudness. Thresholds are shown for the large and the small tonotopic separation of the double pairs (gray bars) and for the respective single pairs (green bars). Each panel contrasts single- with double-pair thresholds when measured at equal overall loudness, i.e., the double pair had lowered levels compared to the respective single pairs such that all stimuli were equally loud. For the large tonotopic separation, the average thresholds were of similar size for the double pair ($BD^{MI}$: 161 μs) and the respective single pairs ($B^{HI}$: 172 μs, $D^{HI}$: 141 μs), whereas for the small tonotopic separation, the aver-age threshold for the double pair ($CD^{MI}$: 234 µs) appeared larger than those of the respective single pairs ($C^{HI}$: 141 µs, $D^{HI}$: 159 µs). Two one-way repeated measures ANOVAs performed on the logarithmically transformed thresholds did not find any significant effect of stimulation type, neither for the large tonotopic separation [$F_{1,9} = 0.08$, $p = 0.7895$] nor for the small tonotopic separation [$F_{1,13} = 4.34$, $p = 0.0575$].

Individual listeners' ITD thresholds as a function of current level (in % DR) are shown for the single pairs in Fig. 5 and for the double pairs in Fig. 6 (small, tinted symbols). Current levels were arithmetically averaged across ears and, for double pairs BD and CD, additionally averaged across electrode pairs. Further, thresholds were grouped by level condition ($L^{HI}$, $L^{MI}$, and $L^{LO}$ for the double pairs BD and CD; $L^{HI}$ and $L^{MI}$ for the single pairs and the double pair DD) by arithmetically averaging the current levels across listeners for each level condition. The levels averaged across ears and listeners in % DR were approximately 54 % (B), 58 % (C), 53 % (D), 55 % (BD), 55 % (CD), and 53 % (DD) in condition $L^{HI}$, 38 % (B), 42 % (C), 38 % (D), 41 % (BD), 39 % (CD), and 38 % (DD) in condition $L^{MI}$, 35 % (BD) and 33 % (CD) in condition $L^{LO}$. The corresponding thresholds, geometrically averaged across listeners within each of the level conditions, are shown as well in Fig. 5 and Fig. 6 (large symbols). The thresholds decreased with increasing current level, irrespective of the tonotopic position or distance. Average thresholds for double pairs tended to increase with decreasing tonotopic distance between the stimulating electrode pairs, with the largest thresholds found for double pair DD. A linear mixed-effects model was fitted to the logarithmically transformed double-pair thresholds with the fixed effects tonotopic separation and current level and a listener-wise intercept as a random effect. The toolbox lme4 (Bates et al. 2014) implemented in R (R Core Team 2013) was used. A type-3 ANOVA was performed to test for significant fixed effects. Denominator degrees of freedom were calculated using the Satterthwaite approximation as implemented in the toolbox lmerTest (Kuznetsova et al. 2014). Main effects of tonotopic separation (the larger the separation, the lower the thresholds) [$F_{2,42.99}$ = 4.80, $p$ = 0.0132] and current level (the higher the current level, the lower the thresholds) [$F_{1,46.43}$ = 36.36, $p$ < 0.0001] were significant, while the interaction was not significant [$F_{2,42.59}$ = 1.37, $p$ = 0.2654]. Post-hoc testing on the factor tonotopic separation (*glht* in the multcomp toolbox, Hothorn et al., 2008; including Bonferroni correction) revealed increased thresholds for the double pair DD as compared to the double pairs BD [$p$ < 0.0001] and CD [$p$ = 0.0003]; thresholds for the double pairs BD and CD were not significantly different from each other [$p$ = 0.1022]. Note that different listeners had slightly different tonotopic inter-electrode distances. A brief, systematic analysis of thresholds as a function of tonotopic inter-electrode distance was performed. No evidence was found that the different inter-electrode distances confounded the effect of tonotopic separation.

## DISCUSSION

*Across-electrode integration*

Our results show that the across-electrode integration of ITD, i.e., the improvement in ITD sensitivity due to the combination of consistent ITD information across two electrode pairs, depended on two contributing factors: tonotopic separation and current level. For a large tonotopic separation and constant levels, stimulation with the double pair led to an improved ITD sensitivity compared to that of the respective single pairs (Fig. 3, left panels), suggesting integration of the ITD information provided by both electrode pairs according to our hypothesis. For a small tonotopic separation and constant levels, ITD sensitivity for single and double pairs was of similar size indicating no benefit of across-electrode ITD integration (Fig. 3, middle panels). When comparing at equal overall loudness, the thresholds for double pairs were similar or even larger than for single pairs, indicating no benefit from across-electrode integration of ITD information for any spacing (Fig. 4).

Ihlefeld et al. (2014) found significantly better ITD sensitivity for a double pair compared to the respective single pairs (see their statistical results from p. 5). In that study, a moderate tonotopic separation (somewhat between our large and small separation) was used. The current levels of the component electrodes were kept constant. However, asymmetries in performance between the two single pairs did not allow to draw conclusions about whether the improved sensitivity for double pairs was due to combining ITD information across pairs or ignoring the worse of the two pairs. In contrast, Francart et al. (2015) compared their results at equal overall *loudness* and found similar performance for single-pair and triple-pair stimulation. However, note that both of the mentioned studies used amplitude-modulated, high-rate pulse trains. Even though their results appear to be compatible with our data, we point out that the sensitivity to ITD conveyed in the ongoing envelope of high-rate pulse trains may differ from the sensitivity to ITD conveyed in unmodulated, low-rate pulse trains.

*Tonotopic separation*

While an increased ITD sensitivity for the double pair compared to that of the respective single pairs was found for the large tonotopic separation (Fig. 3, left panels), the ITD sensitivity was similar for single and double pairs when tested with the small tonotopic separation (Fig. 3, middle panels). This suggests that, in line with our hypothesis, the tonotopic distance between electrode pairs influenced the process of integrating ITD information across electrodes. Accordingly, a significant effect of tonotopic separation on ITD sensitivity of double pairs was found: The larger the tonotopic separation, the better the ITD sensitivity (Fig. 6). Highest thresholds were found for the condition with twice the pulse rate representing infinitely narrow electrode spacing. This is in agreement with previous studies showing impaired ITD sensitivity for increasing pulse rates (e.g., Laback et al. 2007; van Hoesel 2007; van Hoesel et al. 2009). One way of interpreting the deterioration of ITD sensitivity with increasing pulse rate is by considering the refractory effects of the auditory nerve fibers. The reduction of the dead time between subsequent pulses may lead to deteriorated ITD sensitivity due to the reduced excitability of nerve fibers, an effect shown with respect to interaural en-velope delay sensitivity (Francart et al. 2015; Laback et al. 2011). This indicates that, at least for very narrow electrode spacing between two electrode pairs, parts of the stimulated neural population may receive a higher effective pulse rate due to the overlap between the excitation patterns of the two electrodes which, in turn, might interact with the benefit of integrating ITD information across electrodes.

Francart et al. (2015) did not find an effect of electrode separation on sensitivity to ITD conveyed in the ongoing envelope of high-rate pulse trains, which seems contrary to our results showing improved sensitivity with larger tonotopic separation. They used triple pairs with either adjacent electrodes or a tonotopic separation of four electrodes. The difference to our findings might result from the fact that they tested implants with narrower inter-electrode spacings than in the present study. Their "spaced" triple pair corresponds to approximately half the distance in millimeters we used for the small tonotopic separation tested in our study. Consequently, their lack of finding an effect of electrode spacing might be related to the small range of tonotopic separation they had tested.

*Current level*

We found a substantial positive effect of current level on ITD sensitivity, i.e., ITD thresholds decreased with increasing current level, irrespective of stimulation type and tonotopic place or distance (Fig. 5 and Fig. 6). This is consistent with our hypothesis suggesting improved ITD sensitivity with increasing level. The results are in agreement with two listeners tested by van Hoesel (2007) showing better ITD sensitivity for 80 % DR pulse trains than for 60 % DR pulse trains with rates from 100 to 600 pps (the stimulation levels in % DR were determined at 400 pps and held fixed for the other rates). As already suggested by van Hoesel (2007), the improved sensitivity at higher levels might be due to an increased number of neurons providing a more accurate representation of temporal information to the auditory processing pathways evaluating ITDs.

The higher the current level of a stimulus and the larger the number of stimulating electrodes, the higher the total amount of neural activity and thus also the expected loudness (McKay 2004). The current levels of the double pairs had to be decreased in order to be matched in loudness to the respective single pairs. From that it might be presumed that the improved ITD sensitivity for double pairs, which was found for large tonotopic separation and constant current levels, was not due to the integration of ITD information across electrode pairs per se but rather due to the confounding increase in loudness.

On the contrary, higher levels might also negatively affect ITD sensitivity in multiple-electrode stimulation. Increasing the current level causes larger current spread which, in turn, affects the extent of peripheral channel interactions (Franck et al. 2003). Larger current spread might yield a larger overlap between the excitation patterns when stimulating multiple-electrode pairs, especially with small tonotopic separation. This might have a negative impact on performance, as seen for example for some CI listeners in a study investigating their single-electrode discrimination ability as a function of level (Pfingst et al. 1999). If such a negative effect due to channel interactions had existed in our set-up, double-pair thresholds would have remained constant or increased with increasing levels, at least for double pairs with small or zero tonotopic separation. However, lower ITD thresholds were found for higher levels (compare our Fig. 6), suggesting that the positive level effect, i.e., the effect of improved ITD sensitivity with increasing level, dominated the performance.

Our results can be compared with observations made in other discrimination tasks which were also based on temporal information such as pulse-rate discrimination (Pfingst et al. 1994), modulation frequency discrimination (Galvin III et al. 2015), or modulation detection (e.g., Galvin III et al. 2014; Galvin and Fu 2005). The evaluation of single- and multi-channel modulation detection thresholds (MDTs) in CI users showed a significant positive effect of presentation level, i.e., improved performance with increasing levels (Galvin III et al. 2014). Similar to our data, improved

MDTs for multiple-electrode stimulation were only found when single- and multi-channel MDTs were compared at constant levels of the component electrodes, i.e., when the multi-channel stimulus was louder than the single-channel stimulus. Multi-channel MDTs were, however, degraded compared to the average single-channel MDTs when multi-channel loudness summation was compensated for. This is consistent with our conclusion that the positive level effect dominates the ITD performance.

## CONCLUSIONS

In summary, we found that, compared to single-pair stimulation, presenting consistent ITD cues at two electrode pairs resulted in improved ITD sensitivity only if the tonotopic separation was large and if they were stimulated with the same component-electrode levels as the corresponding single pairs. To accommodate the effect of loudness summation in multiple-electrode stimulation, the ITD sensitivity for single- and double-pair stimulation was further evaluated at equal overall loudness, which is clinically more relevant. Presenting ITD information at two electrode pairs did not provide a benefit for any tonotopic spacing.

Overall, our results suggest the contribution, or rather, the counteraction of the factors current level and channel interaction to the process of combining ITD information across electrodes. Provided sufficient stimulus levels and that the stimulating electrode pairs are widely spaced, CI listeners can benefit from combining consistent ITD information across multiple electrodes. Further investigation of the contribution of the factors current level and channel interaction to ITD sensitivity seems to be required for improving spatial hearing with future clinical stimulation strategies.

## ACKNOWLEDGMENTS

We thank all our listeners for their patience and their participation in this study. We are grateful to Stefan-Marcel Pok for help recruiting listeners, Michael Mihocic for help collecting the data, and Peter Schleich and Peter Nopp for fruitful discussions. We thank the Institute of Ion Physics and Applied Physics, Leopold-Franzens-University of Innsbruck, Austria for providing the equipment for direct electric stimulation. We thank the associate editor Ruth Litovsky and two anonymous reviewers for helpful comments on an earlier version of this article. This study was supported by MED-EL Corporation.

**APPENDIX**

For the interested reader, Fig. 7 presents the individual listeners' ITD thresholds for single pairs (open symbols) and double pairs (filled symbols), plotted for large, small, and zero tonotopic separation of the double pairs. Thresholds are shown for the three level conditions $L^{HI}$ (circles), $L^{MI}$ (triangles), and $L^{LO}$ (squares). Colors differentiate between the single pairs B (red), C (green), and D (blue). Note that, for CI52 and CI61, only conditions with small and zero tonotopic separations were tested.

# TABLES

| Listener | Implants (L and R) | Age at testing (yr) | Etiology | Age at onset of deafness (yr) | Age at implantation (yr) | | Duration of bilateral stimulation (yr) |
|---|---|---|---|---|---|---|---|
| | | | | | L | R | |
| CI1 | C40+ | 29 | Meningitis | 13 | 14 | 14 | 15 |
| CI24 | C40+ | 51 | Progressive | 39 | 42 | 43 | 8 |
| CI52 | Pulsar (L) C40+ (R) | 74 | Morbus Menière | 61 | 68 | 63 | 6 |
| CI58 | Concerto | 62 | Progressive | 58 | 62 | 61 | 0.6 |
| CI60 | Pulsar | 66 | Meningitis | 58 | 58 | 58 | 8 |
| CI61 | Concerto | 71 | Progressive | adult | 70 | 69 | 1 |
| CI62 | C40+ | 13 | Connexin 26 disorder | 0 | 3 | 1 | 10 |
| Mean | | 52.3 | | | 45.3 | 44.1 | 6.9 |

TABLE 1: Listener data. Age at onset of deafness refers to the earlier of the two ears and may also relate to the age at the approximate onset of profound hearing loss when no specific date of onset of deafness could be defined (e.g., in case of progressive hearing loss).

| Listener | Ear | Electrode(s) | Phase duration (µs) |
| --- | --- | --- | --- |
| CI52 | Left | 3, 8 | 31.7 |
|  | Left | 11 | 24.6 |
|  | Right | 9 | 33.3 |
| CI58 | Left | 8, 11 | 33.3 |
|  | Right | 11 | 46.7 |

TABLE 2: Deviations from the standard phase duration of 26.7 µs.

| Listener | A (el. num.) L / R | B (el. num.) L / R | C (el. num.) L / R | D (el. num.) L / R | Distance BD (el. num.) L / R | Distance BD (mm) L / R | Distance CD (el. num.) L / R | Distance CD (mm) L / R |
|---|---|---|---|---|---|---|---|---|
| CI1  | 2 / 2 | 5 / 4 | 9 / 7 | 11 / 10 | 6 / 6 | 14.4 / 14.4 | 2 / 3 | 4.8 / 7.2 |
| CI24 | 2 / 3 | 5 / 6 | 8 / 9 | 10 / 12 | 5 / 6 | 12.0 / 14.4 | 2 / 3 | 4.8 / 7.2 |
| CI52 | - / - | 3 / 2 | 8 / 7 | 11 / 9 | n.t. | n.t. | 3 / 2 | 7.2 / 4.8 |
| CI58 | 1 / 1 | 3 / 5 | 8 / 8 | 11 / 11 | 8 / 6 | 16.8 / 14.4 | 3 / 3 | 6.3 / 7.2 |
| CI60 | 1 / 1 | 6 / 6 | 9 / 9 | 11 / 11 | 5 / 5 | 12.0 / 12.0 | 2 / 2 | 4.8 / 4.8 |
| CI61 | 1 / 1 | 5 / 4 | 6 / 6 | 8 / 7 | n.t. | n.t. | 2 / 1 | 4.8 / 2.4 |
| CI62 | 1 / 2 | 4 / 5 | 8 / 8 | 11 / 11 | 7 / 6 | 16.8 / 14.4 | 3 / 3 | 7.2 / 7.2 |
| Mean | 1.3 / 1.7 | 4.4 / 4.6 | 8.0 / 7.7 | 10.4 / 10.1 | 6.2 / 5.8 | 14.4 / 13.9 | 2.4 / 2.4 | 5.7 / 5.8 |

TABLE 3: Single electrode pairs used for binaural stimulation (electrode numbers from apex to base in ascending order). The rightmost columns show the tonotopic distance for the double pairs BD and CD (in number of electrodes and millimeter). n.t. refers to conditions not tested.

# FIGURE CAPTIONS

FIG. 1. Schematic illustration of an example of four pitch-matched electrode pairs labeled from A (the most apical pair) to D (the most basal pair) along the left and right electrode arrays. The best matches were not necessarily equal-numbered electrodes.

FIG. 2. Schematic representation of the used stimuli and task. For both stimulation types, pulse trains are shown for the left (L) and the right (R) ear. Time is shown along the horizontal dimension. In case of double pairs, the temporal offset between the pulses across electrodes is indicated as half the interpulse interval (IPI). The first interval contained the reference stimulus with zero ITD (top row). The second interval contained the target stimulus with non-zero ITD (bottom row).

FIG. 3. ITD thresholds geometrically averaged across listeners for single pairs (green bars) and double pairs (gray bars) representing constant *level*, plotted for large, small, and zero tonotopic separation of the double pairs. Thresholds are shown for the level conditions $L^{HI}$ (top row) and $L^{MI}$ (bottom row). Each panel compares single- and double-pair thresholds for constant levels, i.e., the double pair was louder than the respective single pairs. For the large and small tonotopic separation, the left of the two green bars representing single-pair thresholds depicts the threshold of the more apical single pair. The error bars indicate the geometric standard deviation. Asterisks depict statistically significant differences between thresholds (* $p < 0.05$, **** $p < 0.0001$; two-way repeated measures ANOVA); ns refers to non-significant differences.

FIG. 4. ITD thresholds geometrically averaged across listeners for single pairs (green bars) and double pairs (gray bars) representing equal overall *loudness*. Each panel contrasts single- with double-pair thresholds when measured at equal overall loudness, i.e., all stimuli were equally loud. Other conventions as in Fig. 3.

FIG. 5. Individual listeners' ITD thresholds (small, tinted symbols) and geometric mean thresholds across listeners (large symbols) as a function of current level (in % DR) for single pairs. Current levels were arithmetically averaged across ears. Mean thresholds were grouped by level condition ($L^{HI}$ and $L^{MI}$) by arithmetically averaging the current levels across listeners for each level condition. The error bars indicate the geometric standard deviation.

FIG. 6. Individual listeners' ITD thresholds (small, tinted symbols) and geometric mean thresholds across listeners (large symbols) as a function of current level (in % DR) for double pairs. Current levels were arithmetically averaged across ears and electrode pairs. Other conventions as in Fig. 5.

FIG. 7. Individual listeners' ITD thresholds for single pairs (open symbols) and double pairs (filled symbols), plotted for large, small, and zero tonotopic separation of the double pairs. Thresholds are shown for the three level conditions $L^{HI}$ (circles), $L^{MI}$ (triangles), and $L^{LO}$ (squares). Colors differentiate between the single pairs B (red), C (green), and D (blue). For CI52 and CI61, only conditions with small and zero tonotopic separations were tested.

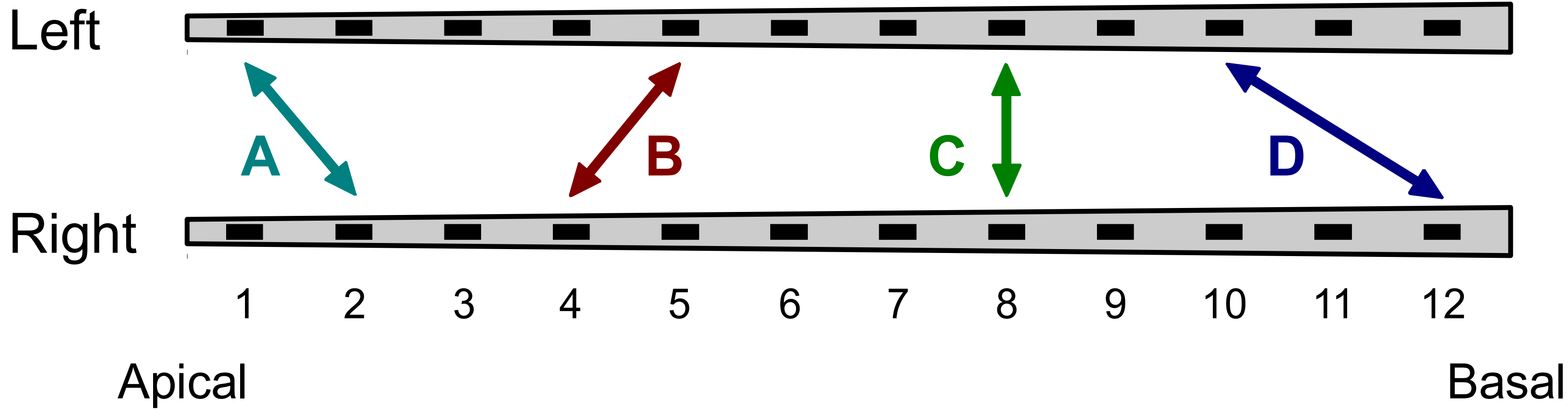

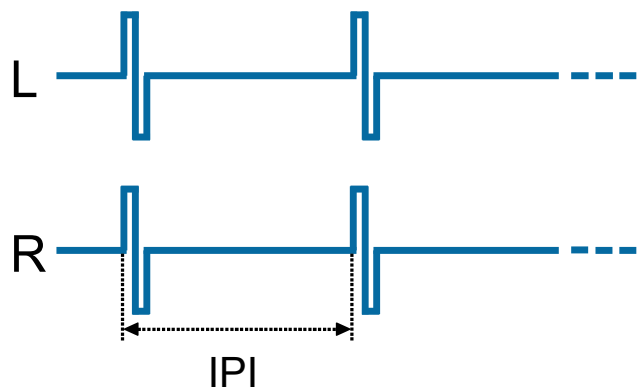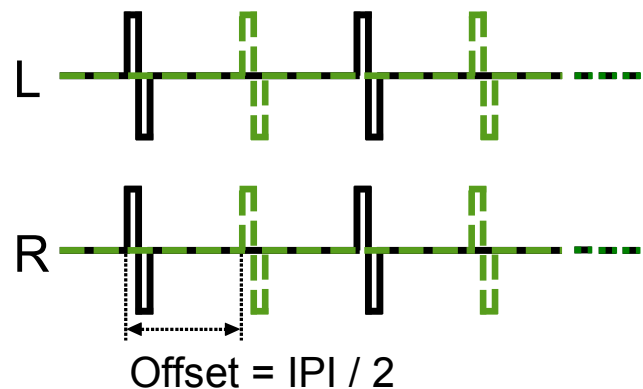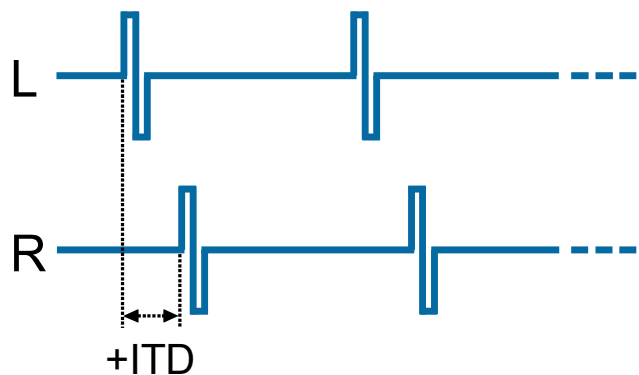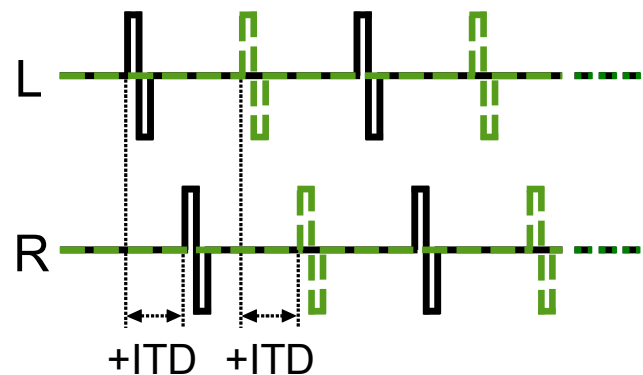

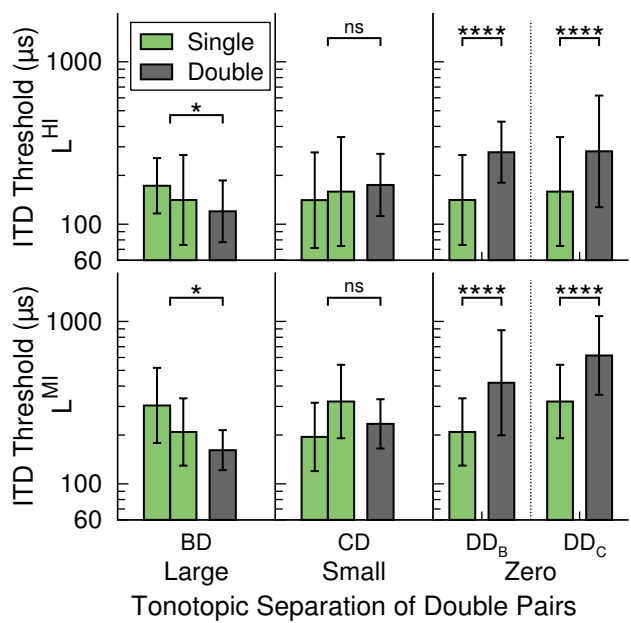

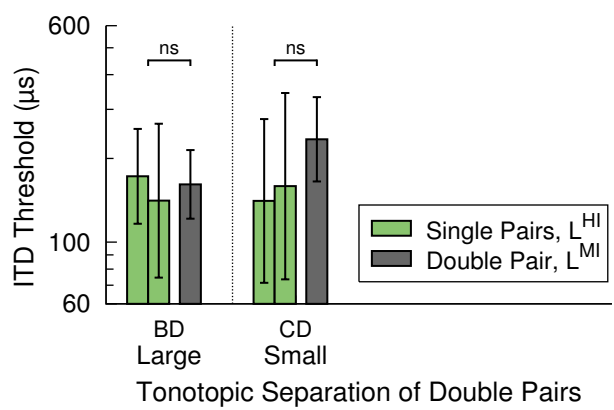

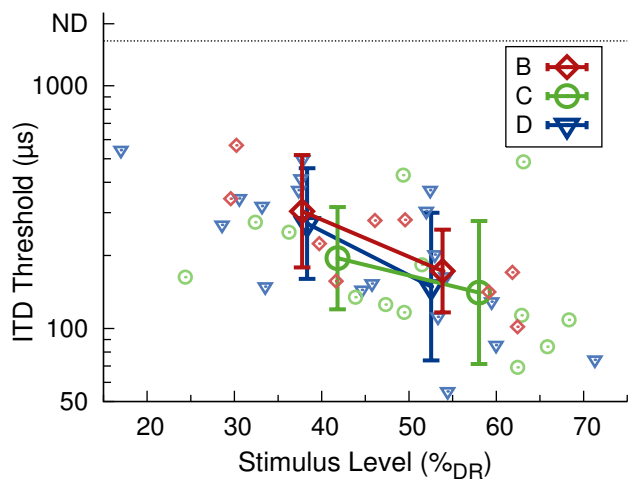

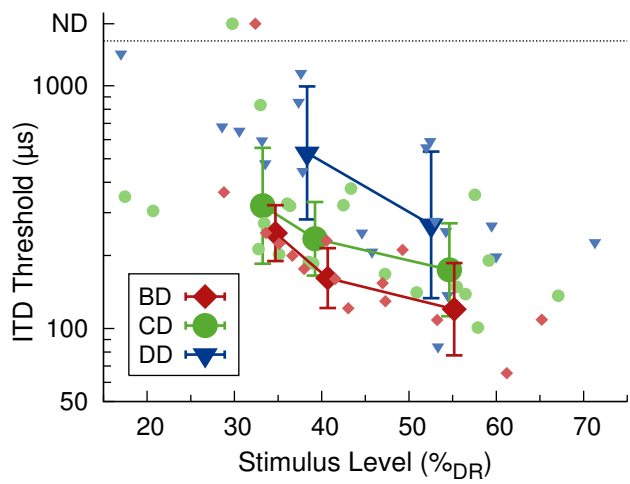

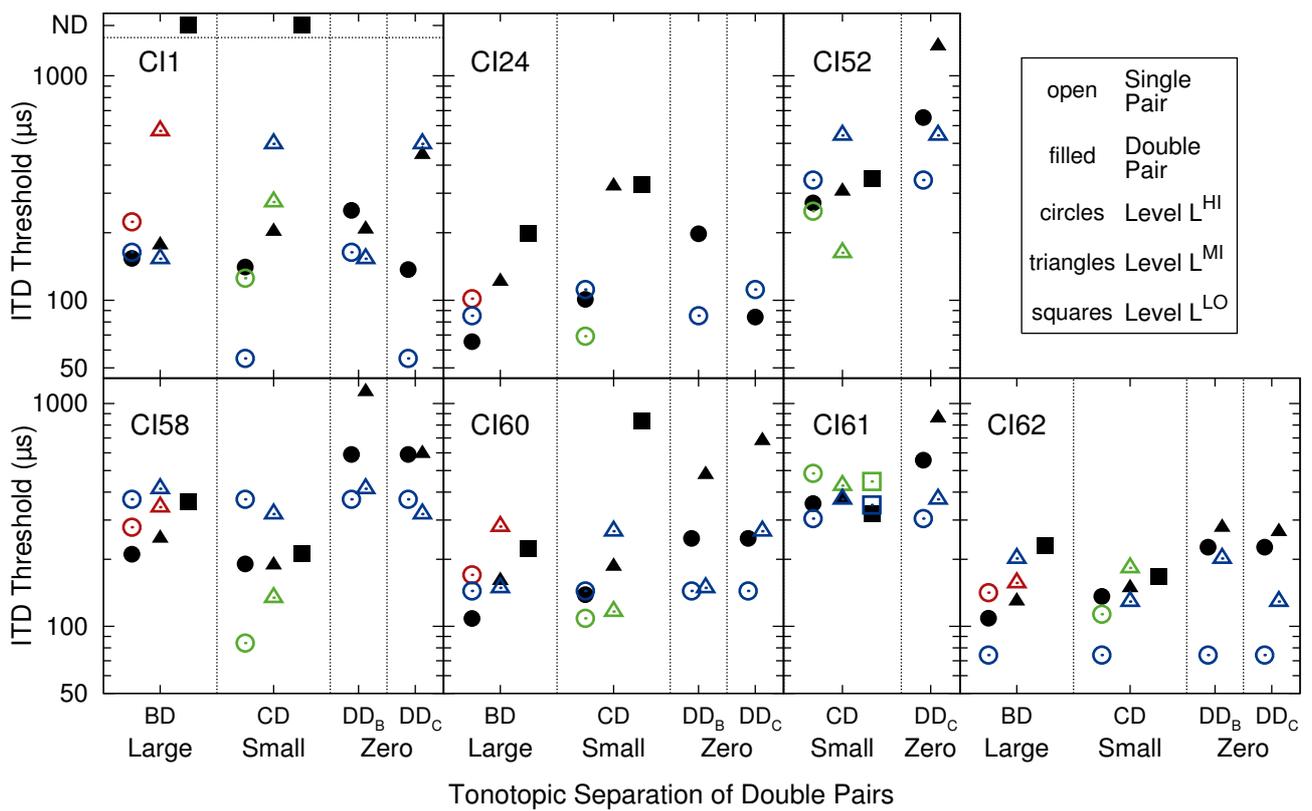